\documentclass[aps,preprintnumbers,amsmath]{revtex4}
\usepackage{amssymb}
\usepackage{graphicx}
\usepackage{dcolumn}
\usepackage{bm}

\begin{document}

\title{Estimation of $\mathcal{P}$-odd correlators in heavy ion
collisions \\ at RHIC energies 62.4 -- 200 GeV}

\author{V.A. Okorokov} \email{VAOkorokov@mephi.ru; okorokov@bnl.gov}
\affiliation{National Research Nuclear University "MEPhI",
Kashirskoe Shosse 31, 115409 Moscow, Russia}

\date{August 17, 2009}

\begin{abstract}
The strength of parity violation effect can be characterized by
correlator value in the framework of local $\mathcal{TIP}$
hypothesis. The energy and centrality dependencies of correlators
for same and opposite charges are discussed for heavy ion
collisions in RHIC energy domain 62.4 -- 200 GeV. The magnetic
field shows a significant increasing at initial energy decreasing
for intermediate and large times. Two possible scenarios for
initial time are investigated in detail. Both scenarios predict
the most close values for initial time in $\mbox{Au+Au}$
collisions at 62.4 GeV. Disagreement between initial time values
obtained for two various scenarios increases with energy
increasing and for collisions of lighter $\mbox{Cu}$-nucleus.
Boundary values for correlators related with possible local
$\mathcal{TIP}$ violation are calculated in the framework of
analytic approach of chiral magnetic effect for $\mbox{Au+Au}$
collisions at 62.4 and 200 GeV. The results are shown as function
of collision centrality. Preliminary experimental STAR data are
compared with predictions of chiral magnetic effect model
directly. Experimental signals consistent with model expectations
for both same and opposite charge correlations in $\mbox{Au+Au}$
collisions at 62.4 GeV at all centralities. At higher energy 200
GeV the model of chiral magnetic effect underestimates same charge
correlator values for central and midcentral events but the same
charge correlations for peripheral events as well as opposite
charge correlations are in the boundaries predicted by model
calculations.


\end{abstract}

\maketitle

\section{\label{intro}Introduction}
Quantum chromodynamics (QCD) as non-Abelian gauge theory contains
non-trivial topological field configurations which deeply relate
with $\mathcal{P/CP}$ invariance of strong interactions. The
non-trivial topology of QCD vacuum opens the possibility for
existence of metastable $\mathcal{P}$ and $\mathcal{CP}$ odd
domains. The possibility of such domains was inferred both from
analysis of an effective chiral theory incorporating axial anomaly
\cite{DKharzeev-PRL-81-512-1998} and from study of space-time
regions occupied by gauge filed configurations with non-trivial
topological charge \cite{DKharzeev-arXiv-9808366,
DKharzeev-NPA-803-227-2008}. These domains can lead to
$\mathcal{P}$ and $\mathcal{CP}$ violation in strong interactions
for some local space region in the vicinity of the deconfinement
phase transition, i.e induce the local strong parity violation
effect. Because topology origin this effect can be assigned also
as local topology induced parity violation
$\left(\mathcal{TIP}\right)$ effect. It was suggested in
\cite{DKharzeev-PRL-81-512-1998} that metastable $\mathcal{P}$ and
$\mathcal{CP}$ odd domains might be created in heavy ion
collisions at high energies. The mechanism by which such domains
can demonstrate themselves, in particular, via separate electric
charges in the presence of a background strong magnetic field --
the chiral magnetic effect -- was suggested in
\cite{DKharzeev-NPA-803-227-2008}. The effect predicts the
preferential emission of charged particles along the direction of
system's angular momentum in the case of the noncentral heavy-ion
collisions due to the presence of nonzero chirality. Since
separation of charge is $\mathcal{CP}$ odd, any experimental
observations of the chiral magnetic effect could be provided a
clear demonstration of the non-trivial topological structure of
the QCD vacuum. Moreover the key requires for the possible local
$\mathcal{TIP}$ violation in strong interactions are the
deconfinement state of matter with restored chiral symmetry
\cite{DKharzeev-NPA-803-227-2008,Fukushima-PRD-78-074033-2008}.
The former is needed to separate (anti-)quarks with opposite
electric charges by a distance larger than nucleon size. The
spatially restored chiral symmetry is required because charge
separation is possible at conserved chirality only. Thus the
positive and reliable experimental results for $\mathcal{P}$ and
$\mathcal{CP}$ violation in the strong interactions would be prove
the clear evidence of deconfinement and chirally symmetric phase
creation and establish experimentally the existence of non-trivial
topological field configurations in QCD and their role in chiral
symmetry breaking
\cite{DKharzeev-private,VAOkorokov-arXiv-0809.3130}. Possible
experimental signals of local parity violation in relativistic
heavy ion collisions were suggested in
\cite{DKharzeev-PRD-61-111901-2000,Voloshin-PRC-62-044901-2000,
Finch-PRC-65-014908-2001, DKharzeev-PLB-633-260-2006}. It was
suggested that signals of local $\mathcal{TIP}$ violation in heavy
ion collisions can be observed through the e-by-e charge
assymetries with respect to the reaction plane. The energy
dependence of strength of the local $\mathcal{TIP}$ violation is
important challenge for study of various phases and critical point
for strongly interacting matter.
\section{\label{sec2}Method}
The averaged correlators for chiral magnetic effect approach are
defined as \cite{DKharzeev-NPA-803-227-2008}
\begin{eqnarray}
\label{Th-corr} a_{\pm \pm}=\frac{\textstyle 1}{\textstyle
N_{\pm}^{2}}\frac{\textstyle \pi^{2}}{\textstyle
16}g\left(b/R,\lambda/R\right)\Phi\left(b/R,\tau,\eta\right),
~~~a_{+-}=-\frac{\textstyle 1}{\textstyle
N_{+}N_{-}}\frac{\textstyle \pi^{2}}{\textstyle
16}h\left(b/R,\lambda/R\right)\Phi\left(b/R,\tau,\eta\right).
\end{eqnarray}
for same and opposite charges respectively, where $N_{\pm}$
denotes the total number of charged particles in the corresponding
pseudorapidity interval. The function
$\Phi\left(b/R,\tau,\eta\right)$ is following
\begin{eqnarray}
\label{Phi-func}\Phi\left(b/R,\tau,\eta\right)=4\kappa\alpha_{s}R^{2}\left[\smash[b]{\sum_{j}q_{j}^{2}}\right]^{2}
\int\limits_{\tau_{i}}^{\tau_{f}}d\tau\tau\left[eB\left(\tau,\eta\right)\right]^{2}.
\end{eqnarray}
Here the proper time $\tau=\sqrt{t^{2}-z^{2}}$ and space-time
rapidity $\eta=0.5\ln\left[(t+z)/(t-z)\right]$. The time integral
on magnetic field is from initial time $\tau_{i}$ to a final time
$\tau_{f}$. The
$g\left(b/R,\lambda/R\right),~h\left(b/R,\lambda/R\right)$ are
some universal functions which depend on centrality $(b/R)$ and
screening length $(\lambda)$, which is assumed a constant in time
range for integral. Here $\alpha_{s}$ is strong coupling constant,
$R$ - nucleus radius, $\kappa \sim 1$ is a some constant
coefficient, $q_{j}$ is the charge in units of $e$ of a quark with
flavor $j$. This approach is valid for a constant homogeneous
magnetic field. In the overlap region the magnetic field is to a
good degree homogeneous around zero space–time rapidity especially
for large impact parameters. The more detail description is in
\cite{DKharzeev-NPA-803-227-2008}.

In the energy domain under study the Lorentz contraction factor
$\gamma_{L}$ of colliding nuclei is in 31.2 -- 100 range. Hence
the nuclei are Lorentz contracted in the $z$-direction to about 3
(1) percent of their original size for collision energy 62.4 (200)
GeV. Therefore the pancake shape seems valid approximation for two
colliding nuclei. One can assume $\tau_{f}$ is infinity in the
(\ref{Phi-func}) because of dependence of magnetic field absolute
value on $\tau$ shows the rapid decreasing of $\left(eB\right)$
with $\tau$ increasing \cite{DKharzeev-NPA-803-227-2008}. Analytic
approach for magnetic field from \cite{DKharzeev-NPA-803-227-2008}
allows us to redefine the (\ref{Phi-func}) and derive the
following analytic formula:
\begin{eqnarray}
\label{Phi-func-Redef} \Phi\left(b/R,\tau_{i},Y_{0}\right)=
4\kappa\alpha_{s}\left(Z\alpha_{\mbox{\scriptsize{EM}}}\right)^{2}\left[\smash[b]{\sum_{j}q_{j}^{2}}\right]^{2}\frac{\textstyle
R}{\textstyle \tau_{i}}~
\mbox{e}^{-Y_{0}}\left[c^{2}f^{2}\left(\frac{\textstyle
b}{\textstyle R}\right)+\frac{\textstyle 32}{\textstyle
5}cf\left(\frac{\textstyle b}{\textstyle R}\right)\frac{\textstyle
b\sqrt{R}}{\textstyle
\tau_{i}^{3/2}}~\mbox{e}^{-3Y_{0}/2}+16\frac{\textstyle
b^{2}R}{\textstyle \tau_{i}^{3}}~\mbox{e}^{-3Y_{0}}\right],
\end{eqnarray}
where $c \simeq 0.599$, $\alpha_{\mbox{\scriptsize{EM}}}$ denotes
the electromagnetic fine structure constant, $f(b/R)$ is some
universal function which depends on centrality, and $Y_{0}$ -- the
beam rapidity. The analytic approach for magnetic field and, as
consequence, formula (\ref{Phi-func-Redef}) are valid for
$R/\sinh\left(Y_{0}\right) \lesssim \tau \lesssim R$. Thus
correlator values depend on $\tau_{i}$ choice, beam
characteristics (ion type and initial energy), and centrality in
the 2D approach with sharp boundary of surface for density of
incoming nuclei.

Experimental observation of charge separation effect is possible
only by correlation techniques because of direction of charge
separation may change event by event due to random sign of the
topological charge of the local domain. The $\mathcal{P}$-even
experimental observable which is sensitive to the charge
separation relates with averaged correlator from chiral magnetic
effect model as following \cite{Voloshin-PRC-70-057901-2004}:
\begin{eqnarray}
\label{Exp-corr}
\langle\cos\left(\phi_{\alpha}+\phi_{\beta}-2\Psi_{RP}\right)\rangle-\left[B_{in}-B_{out}\right]-
\langle v_{1}^{\alpha}v_{1}^{\beta}\rangle=
-\left<a_{\alpha}a_{\beta}\right>,
\end{eqnarray}
where $\alpha,\beta=-,+$ denote the particle electric charges,
$\Psi_{RP}$ -- azimuthal angle of the reaction plane, $v_{1}$ --
directed flow parameter. The average in (\ref{Exp-corr}) is taken
over all pairs in some event and then over all events with given
centrality. The experimentally measured correlator
$\langle\cos\left(\phi_{\alpha}+\phi_{\beta}-2\Psi_{RP}\right)\rangle$
represents the difference between correlations projected onto an
axis in the reaction plane and the correlations projected onto an
axis perpendicular to the reaction plane. The important advantage
of using (\ref{Exp-corr}) is that it removes all the correlations
among particles $\alpha$ and $\beta$ that are not related to the
reaction plane orientation
\cite{Borghini-PRC-66-014905-2002,STAR-PRL-92-062301-2004}. The
second term on the left side is the difference between background
contribution of the in-plane correlations $B_{in}$ and background
contribution of the out-of-plane correlations $B_{out}$. Note that
the contribution given by the term $\langle
v_{1}^{\alpha}v_{1}^{\beta}\rangle$ be neglected because directed
flow averages to zero in a rapidity region symmetric with respect
to midrapidity region which is considered in this paper
\cite{STAR-PRL-92-062301-2004,Voloshin-0907.2213}.

\section{\label{sec3}Results}
\begin{figure}
\begin{minipage}[b]{.48\linewidth}
\centering{\includegraphics[width=9.0cm,height=9.0cm]{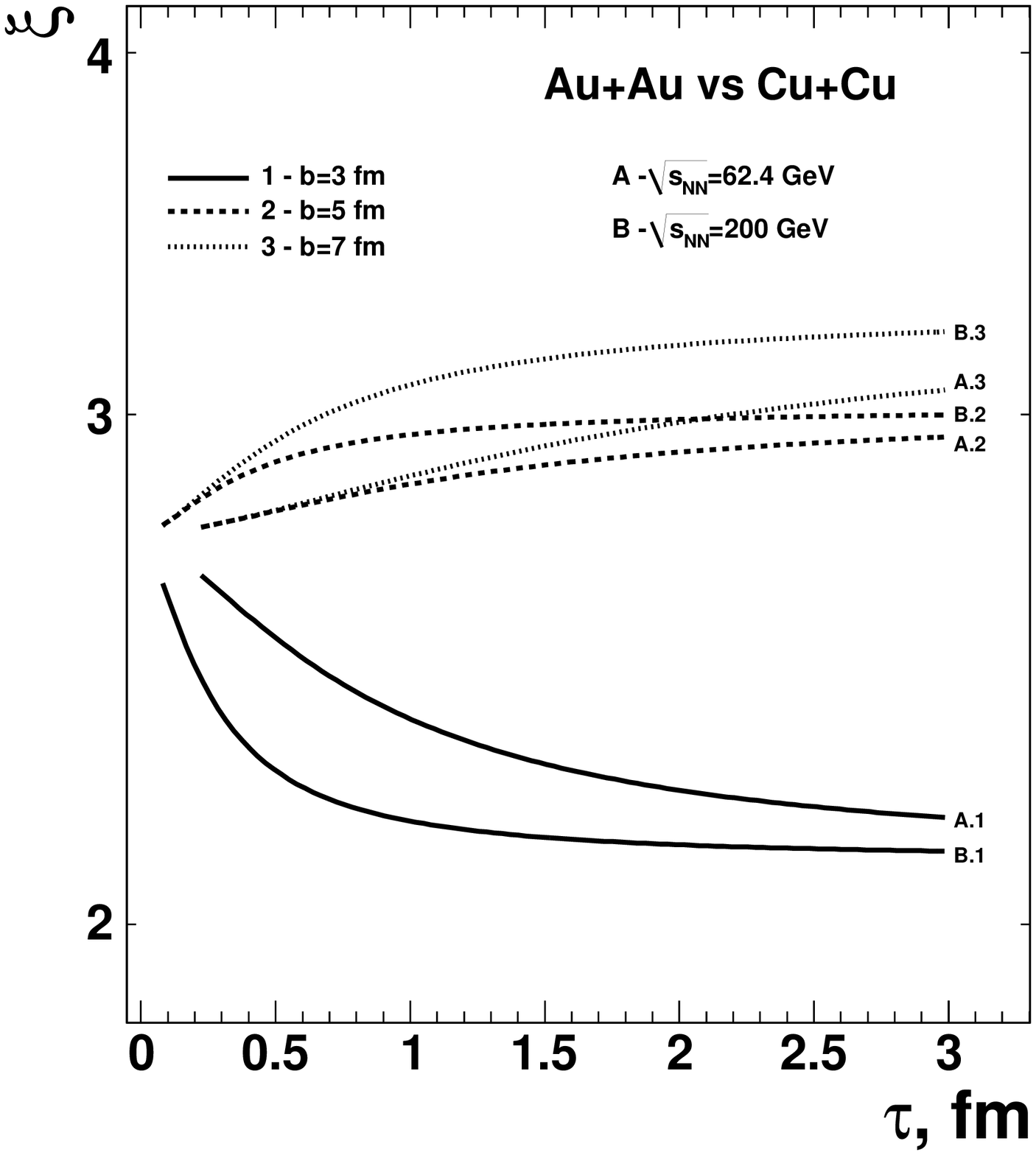}}
\caption{Magnetic field ration depends on $\tau$ for two various
RHIC beams at equal energies and for various impact parameters in
$R/\sinh\left(Y_{0}\right) \lesssim \tau$
domain.}\label{fig:MagFieldRatio-DiffBeamA}
\end{minipage}\hfill
\begin{minipage}[b]{.48\linewidth}
\centering{\includegraphics[width=9.0cm,height=9.0cm]{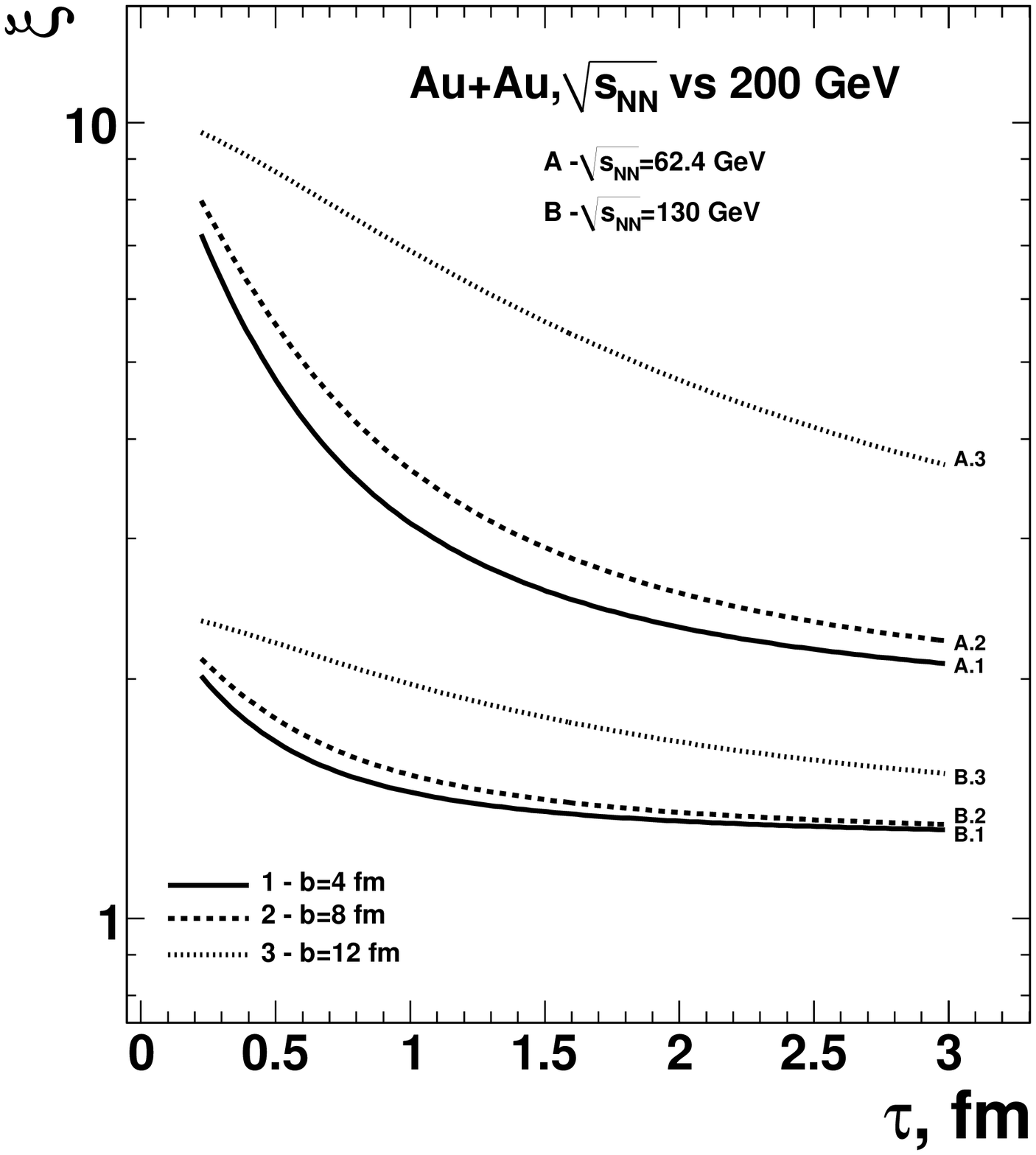}}
\caption{Magnetic field ratio for $\mbox{Au}$-beam with energy
$\sqrt{s_{NN}}$ to the same beam type with 200 GeV energy depends
on $\tau$ and for different $b$ values in
$R/\sinh\left(Y_{0}\right) \lesssim \tau$
domain.}\label{fig:MagFieldRatio-DiffBeamEnergy}
\end{minipage}
\end{figure}

We consider the magnetic fields created in non-central heavy-ion
collisions for two different beams in the framework of analytic
approach of chiral magnetic effect at $R/\sinh\left(Y_{0}\right)
\lesssim \tau$ \cite{DKharzeev-NPA-803-227-2008}. The ratio of
magnetic fields created in various collisions,
$\xi=\left(eB\right)_{1} / \left(eB\right)_{2}$ is investigated
here in dependence on proper time.
Fig.\ref{fig:MagFieldRatio-DiffBeamA} shows the $\xi$ for two
different RHIC beam types at equal collision energies and the
ratio of strengths of magnetic fields created by $\mbox{Au}$ beams
at various energies is presented at
Fig.\ref{fig:MagFieldRatio-DiffBeamEnergy}. Magnetic field is
larger for more heavy nucleus at any energies as expected and
magnetic field value is more sensitive to beam energy than beam
ion type. First estimates for $\xi$ at low energies $\sqrt{s_{NN}}
\! < \! 20$ GeV were obtained also in pancake approach with sharp
surface boundary \cite{VAOkorokov-NICA-RT3}. In this analytic
approach the magnetic field at intermediate energies is larger
significantly than that for high energy domain. It should be
stressed additionally that 2D pancake picture is rough approach
for low energy domain. Thus the magnetic field value decreases
more rapidly for higher energies at intermediate and large $\tau$.
The recent UrQMD calculations \cite{Skokov-arXiv-0907.1396} show
magnitude of magnetic field $\left(eB\right)_{y} \sim m_{\pi}^{2}$
for $\mbox{Au+Au}$ collisions at at highest RHIC energy
$\sqrt{s_{NN}}=200$ GeV. This estimation is similar to the
magnitude calculated in the framework of chiral magnetic effect
model. The UrQMD model as well as numerical estimations for chiral
magnetic effect shows the increasing of strength of background
magnetic field with initial energy increasing at very small times
$R/\sinh\left(Y_{0}\right) \gtrsim \tau$.

\begin{figure*}
\includegraphics[width=15.0cm,height=10.5cm]{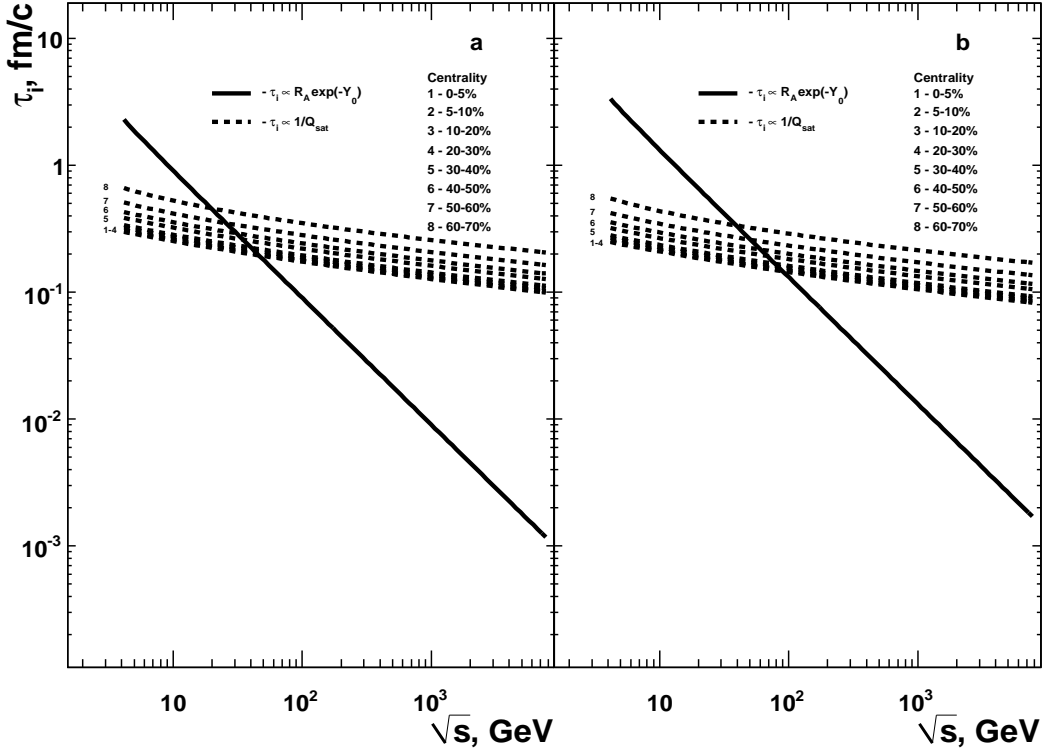}
\caption{Energy dependence of initial time at two scenarios for
$\mbox{Cu}$ (a) and $\mbox{Au}$ (b) beam.} \label{fig:tau-Energy}
\end{figure*}

As seen from (\ref{Phi-func-Redef}) expressions for variations of
the difference between the number of raising and lowering
transitions depend on the choice of the initial time
\cite{DKharzeev-NPA-803-227-2008}. In the framework of analytic
approach for chiral magnetic effect two possible scenarios for
initial time definition were suggested. If the time at which the
topological charge changing processes is less than the Lorentz
contracted size of the system, one can use the Lorentz contracted
size scale for estimation of up limit of initial time:
$\tau_{i}^{I}=\zeta R\exp(-Y_{0})$, where $\zeta \simeq 2$. On the
other hand, if the time at which the topological charge changing
processes is larger than the Lorentz contracted size of the
system, we should use the following scale for estimation of
initial time: $\tau_{i}^{II} \sim 1/Q_{\mbox{\scriptsize{sat}}}$,
here $Q_{\mbox{\scriptsize{sat}}}$ is the saturation scale
\cite{DKharzeev-NPA-803-227-2008}. As seen the $\tau_{i}^{II}$
depends on centrality and beam type due to
$Q_{\mbox{\scriptsize{sat}}}$ \cite{DKharzeev-PLB-507-121-2001}.
These scenarios define the boundary values for initial time. Thus
correlator values calculated for $\tau_{i}^{I}$ and
$\tau_{i}^{II}$ are limit values for certain kinematics and beam
type. Fig.\ref{fig:tau-Energy} shows energy dependence of initial
time at various scenarios for $\mbox{Cu}$
(Fig.\ref{fig:tau-Energy}a) and for $\mbox{Au}$
(Fig.\ref{fig:tau-Energy}b) beams.

\begin{figure}[h]
\resizebox{0.5\textwidth}{!}{%
\includegraphics[width=9.0cm,height=9.0cm]{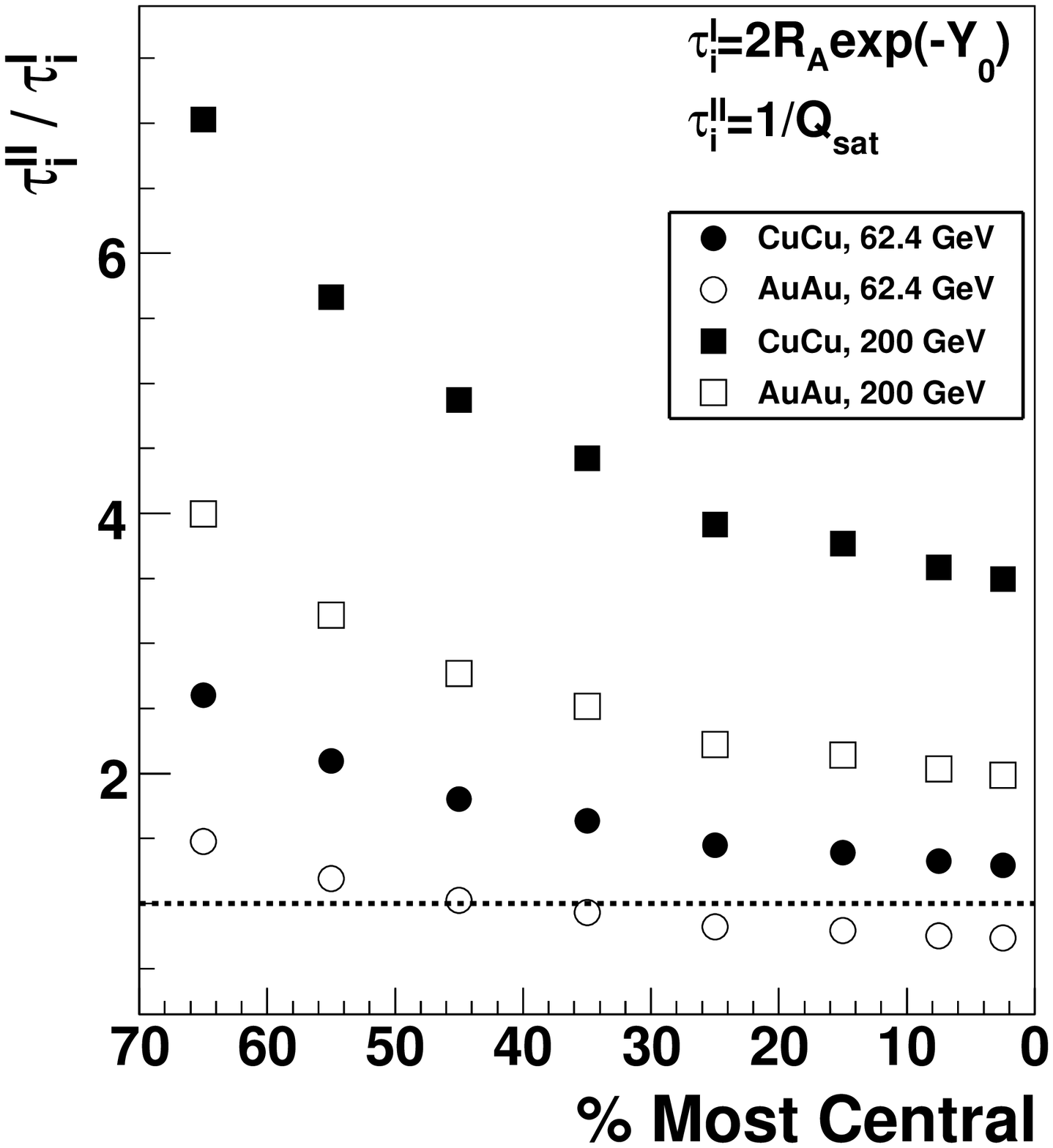}
} \caption{Ratio of initial time values for two scenarios depends
on centrality for two RHIC energies. Dashed line is at unity.}
\label{fig:tau-Ratio}
\end{figure}

Ratio of initial time values derived from scenarios under study is
demonstrated on the Fig.\ref{fig:tau-Ratio} for two RHIC energies
and various beam types. One needs to emphasize that both scenarios
for initial time show the similar values up to order of magnitude,
at least, for RHIC energy domain 62.4 - 200 GeV. Two scenarios
agree well especially at 62.4 GeV for central and midcentral
collisions and all nuclei under study. The differences between
these scenarios are significantly larger for lower and higher
collision energies. Therefore one can expect wider allowed domain
for correlator values at lower and higher initial energies than
that for RHIC ones.

Thus it should be stressed the unique situation for RHIC energy
domain 62.4 - 200 GeV: (i) the pancake shape is a really good
approximation and one can consider 2D picture for magnetic field
calculation; (ii) the two various scenarios for initial time
result in a similar estimations.

The important feature of (\ref{Phi-func-Redef}) is the
dependencies of $\Phi$ function (and, as consequence, correlators
too) on beam type and centrality only for choice of the first
scenario for $\tau_{i}$. Thus if the same colliding nuclei under
study one can obtain the simple relations for correlators at two
various initial energies $\sqrt{s_{1}}$ and $\sqrt{s_{2}}$:
\begin{eqnarray}
\label{Th-corr-Ratio} \frac{\textstyle \left.a_{\pm
\pm}\right|_{\sqrt{s_{1}}}}{\textstyle \left.a_{\pm
\pm}\right|_{\sqrt{s_{2}}}}=\left[\frac{\textstyle \left.
N_{\pm}\right|_{\sqrt{s_{2}}}}{\textstyle \left.
N_{\pm}\right|_{\sqrt{s_{1}}}}\right]^{2},~~~\frac{\textstyle
\left.a_{+-}\right|_{\sqrt{s_{1}}}}{\textstyle
\left.a_{+-}\right|_{\sqrt{s_{2}}}}=\left[\frac{\textstyle \left.
N_{+}N_{-}\right|_{\sqrt{s_{2}}}}{\textstyle \left.
N_{+}N_{-}\right|_{\sqrt{s_{1}}}}\right].
\end{eqnarray}

Based on the (\ref{Th-corr-Ratio}) and energy dependence of
charged particle multiplicity in heavy ion collisions
\cite{DKharzeev-PLB-507-121-2001} one can obtain
\begin{eqnarray}
\nonumber \frac{\textstyle \left.a_{\pm
\pm}\right|_{200}}{\textstyle \left.a_{\pm
\pm}\right|_{62.4}}=\frac{\textstyle
\left.a_{+-}\right|_{200}}{\textstyle \left.a_{+-}\right|_{62.4}}
\simeq 0.527.
\end{eqnarray}

\begin{figure*}
\includegraphics[width=16.0cm,height=16.0cm]{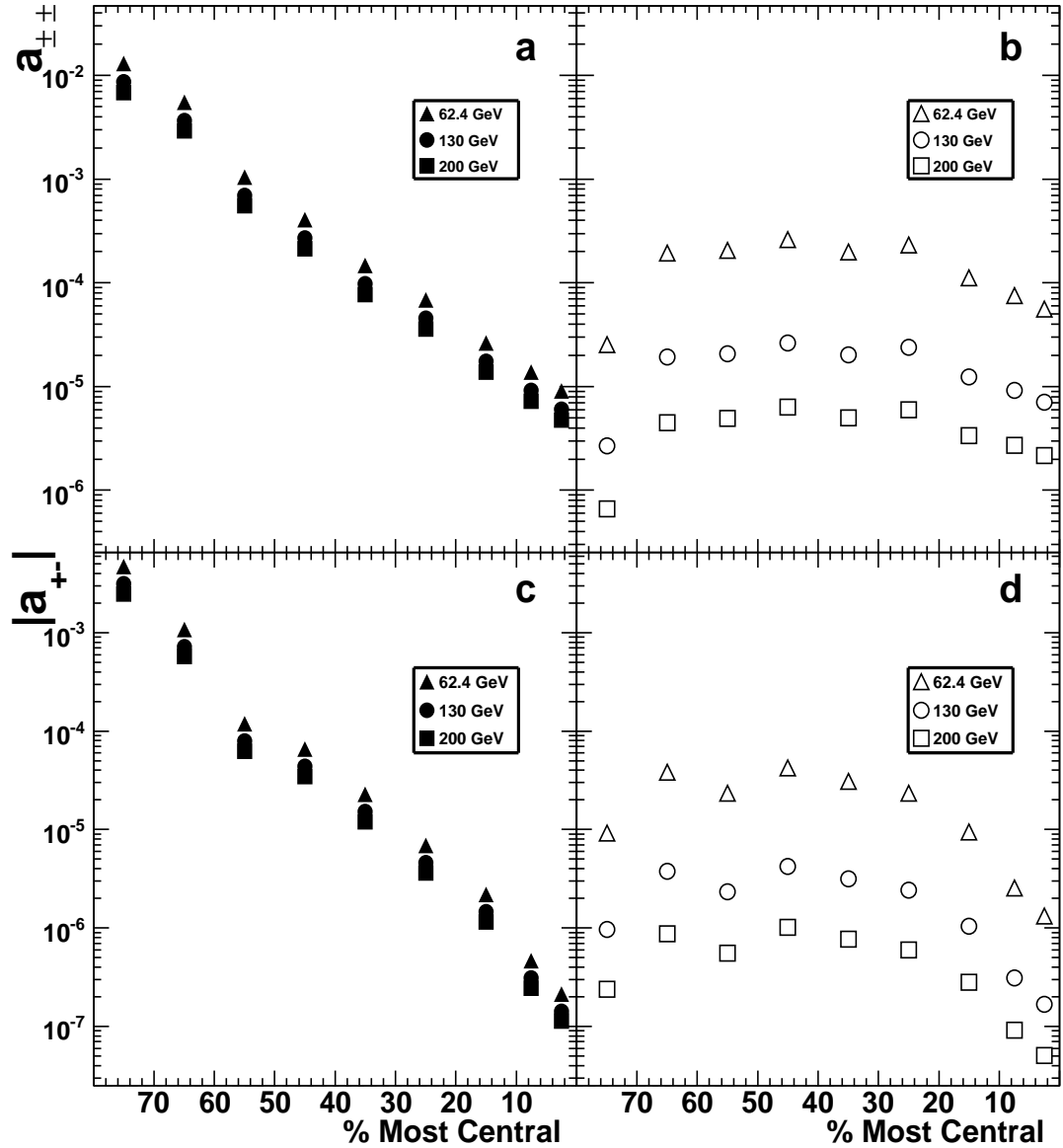}
\caption{Centrality dependence of correlator absolute values for
same (a,b) and opposite (c,d) charges in $\mbox{Au+Au}$ collisions
at RHIC energy domain 62.4 -- 200 GeV in the framework of Lorentz
contracted (a,c) and saturation (b,d) scale scenario for initial
time. Data for 130 GeV are from
\cite{DKharzeev-NPA-803-227-2008}.} \label{fig:Th-MyCorr}
\end{figure*}

\begin{figure}
\begin{minipage}[b]{.48\linewidth}
\centering{\includegraphics[width=9.0cm,height=9.0cm]{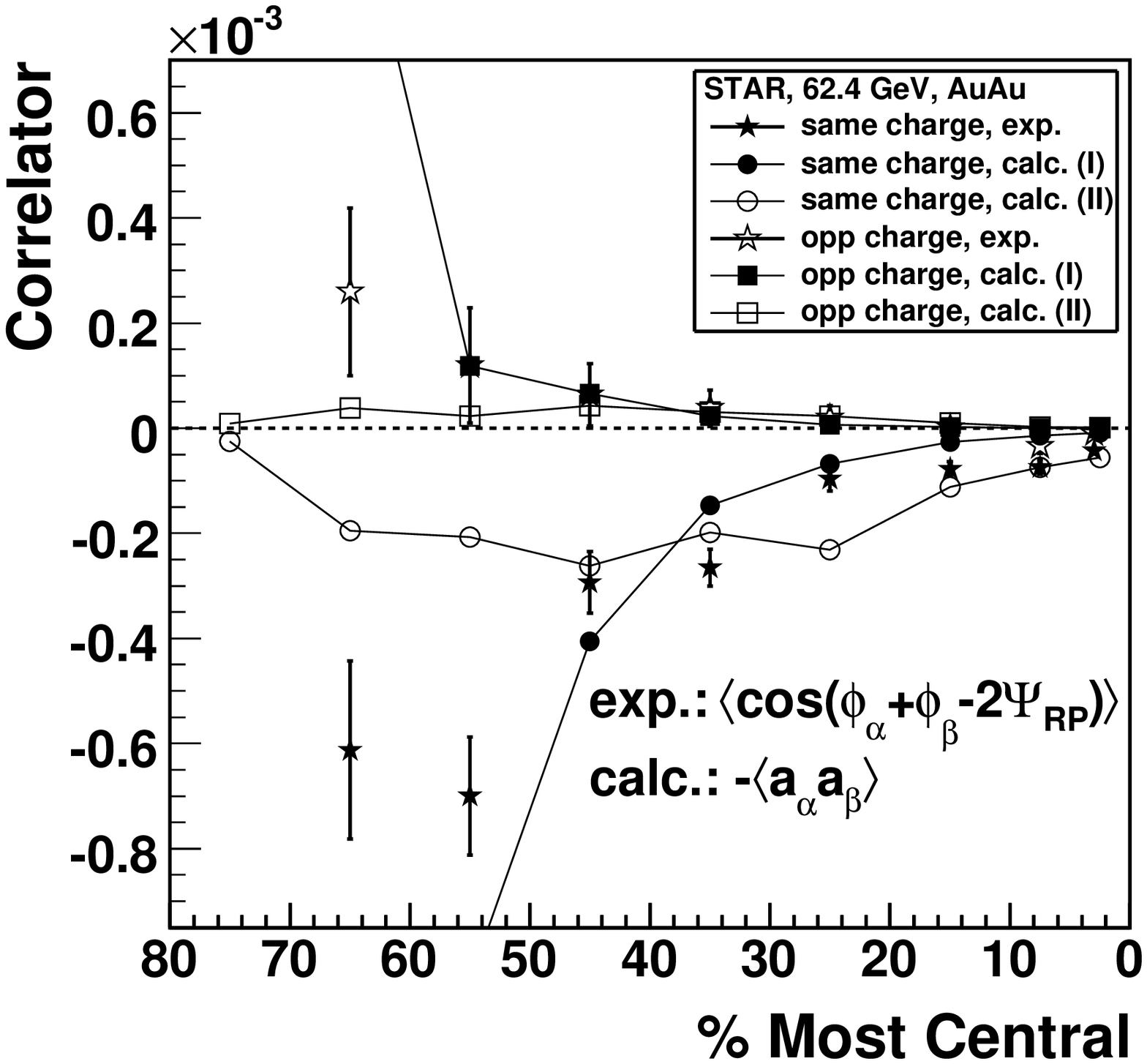}}
\caption{Centrality dependence for correlators calculated in the
framework of analytic approach for chiral magnetic effect at two
initial time scenarios and for preliminary experimental results
obtained by STAR for charged particles in $\mbox{Au+Au}$
collisions at 62.4 GeV
\cite{Voloshin-0907.2213}.}\label{fig:ExpVsTh-AuAu62GeV}
\end{minipage}\hfill
\begin{minipage}[b]{.48\linewidth}
\centering{\includegraphics[width=9.0cm,height=9.0cm]{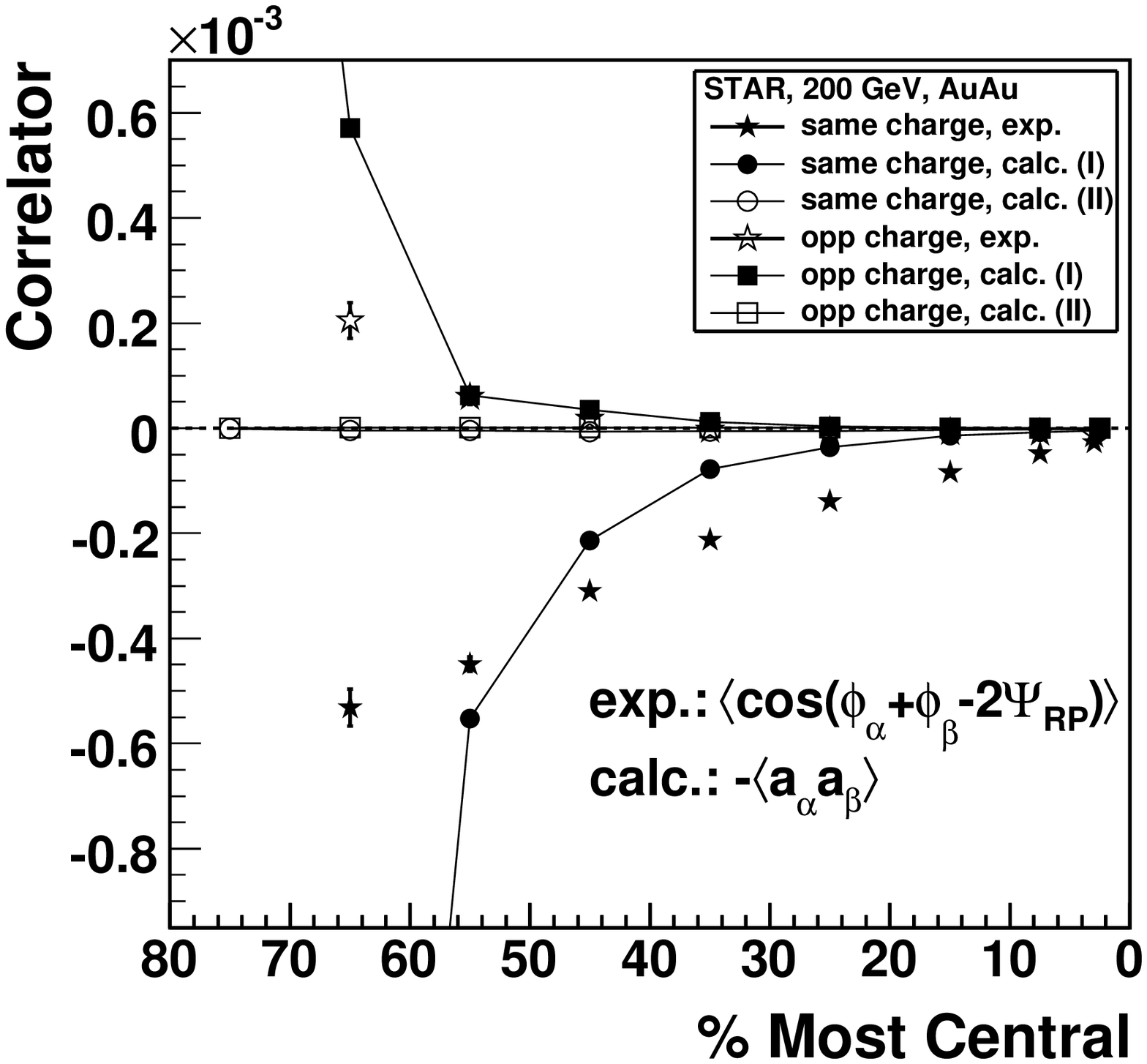}}
\caption{The correlators calculated in the framework of analytic
approach for chiral magnetic effect at two initial time scenarios
and preliminary experimental values obtained by STAR for charged
particles in $\mbox{Au+Au}$ collisions at 200 GeV
\cite{Voloshin-0907.2213} depend on collision
centrality.}\label{fig:ExpVsTh-AuAu200GeV}
\end{minipage}
\end{figure}

The estimations for correlators $a_{\pm}$ and
$\left|a_{+-}\right|$ for $\mbox{Au+Au}$ collisions at
$\sqrt{s_{NN}}=130$ GeV were calculated in
\cite{DKharzeev-NPA-803-227-2008}. The centrality dependence of
correlator values for $\mbox{Au+Au}$ at 62.4 GeV and 200 GeV are
shown at Fig.\ref{fig:Th-MyCorr} for same (a,b) and opposite (b,d)
charge combinations for Lorentz contracted scale (a,c) and
saturation scale (b,d) scenarios for initial time. We have assumed
$N_{+} \simeq N_{-} \simeq N_{ch}/2$ and $\lambda=1$ fm
\cite{DKharzeev-NPA-803-227-2008}, where $N_{ch}$ is the charged
particle multiplicity at midrapidity for centrality bin under
consideration. As expected above both correlators show a fast
increasing at decreasing of collision centrality for
$\tau_{i}^{I}$ scenario (Fig.\ref{fig:Th-MyCorr}a,c). Correlator
estimations in the framework of saturation scale scenario show a
slow increasing at decreasing of collision centrality and
decreasing for most peripheral collisions both for same
(Fig.\ref{fig:Th-MyCorr}b) and opposite (Fig.\ref{fig:Th-MyCorr}d)
charge combinations. But it should be stressed that saturation
scale calculations based on the approach from
\cite{DKharzeev-PLB-507-121-2001} and available data allow us to
make an approximate and rough estimations for
$Q_{\mbox{\scriptsize{sat}}}$ for two most peripheral centrality
bins. Saturation scenario for $\tau_{i}$ results in significantly
smaller values both for same and opposite charge correlators for
midcentral and peripheral events than correlator estimations at
$\tau_{i}^{I}$. It seems the decreasing of correlators for most
peripheral bin is more reasonable than that behaviour in the
framework of the first scenario for $\tau_{i}$ because the
deconfinement matter is created in (mid-)central collisions. Thus
the analytic approach for observables related with possible local
$\mathcal{TIP}$ violation allows us to get limit values for same
and opposite charge correlators.
Direct comparison of expectation values for correlators calculated
in the framework of local $\mathcal{TIP}$ hypothesis and
preliminary experimental STAR data \cite{Voloshin-0907.2213} are
shown at Fig.\ref{fig:ExpVsTh-AuAu62GeV} and
Fig.\ref{fig:ExpVsTh-AuAu200GeV} for collision energies 62.4 GeV
and 200 GeV, respectively. The contribution given by the term from
directed flow in (\ref{Exp-corr}) is assumed equal zero for
experimental data \cite{Voloshin-0907.2213}. Preliminary STAR data
are in the ranges predicted by analytic approach for chiral
magnetic effect for opposite charge correlations at overall
centralities and both energies under consideration. Model values
calculated for the $\tau_{i}^{I}$ scenario are very close to the
preliminary experimental results for opposite charge combinations.
In contrast, the same charge experimental correlations are within
model boundaries for 62.4 GeV at most of centrality bins and for
peripheral bins for 200 GeV. But the model predicts very wide
allowed range in the last case. Thus agreement between
experimental data and model estimations is better for lower
energy.

Perhaps disagreement between preliminary experimental data and
model estimations at 200 GeV for central and midcentral collisions
can be explained by significant difference between background
contributions of in-plane correlations and out-of-plane
correlations: power of jet quenching, for example, depends on
orientation with respect to the reaction plane
\cite{STAR-PRL-93-252301-2004,VAOkorokov-YF-72-155-2009}. This can
leads to the increasing of $\Delta B=B_{in}-B_{out}$. One can
expect that the difference between contributions of jet-like
in-plane correlations and out-of-plane correlations decreases with
energy decreasing. Thus the disbalance $\Delta B$ is lower for
lower energy. On the other hand the exception of the very small
times $R/\sinh\left(Y_{0}\right) \gtrsim \tau$ in present analytic
approach may be one of the reasons for decreasing of calculated
correlator estimations. It seems the influence of this exception
is more significant at 200 GeV energy namely than at 62.4 GeV
because larger magnitude at very small times and faster decreasing
are expected for magnetic field at higher energy.

In additional one can get a rough estimations for
$\mathcal{P}$-odd correlators for first initial time scenario at
LHC energy $\sqrt{s_{NN}}=5.5$ TeV due to
$R_{\mbox{\scriptsize{Au}}} \simeq R_{\mbox{\scriptsize{Pb}}}$.
Thus
\begin{eqnarray}
\nonumber \frac{\textstyle \left.a_{\pm
\pm}\right|_{\mbox{\scriptsize{LHC}}}}{\textstyle \left.a_{\pm
\pm}\right|_{62.4}}=\frac{\textstyle
\left.a_{+-}\right|_{\mbox{\scriptsize{LHC}}}}{\textstyle
\left.a_{+-}\right|_{62.4}} \simeq 0.085
\end{eqnarray}
and interval estimations for same and opposite charge correlators
are (from central to peripheral centrality bins) $\left.a_{\pm
\pm}\right|_{\mbox{\scriptsize{LHC}}} \sim
\left[10^{-6}-10^{-5}\right]$ and
$\left.a_{+-}\right|_{\mbox{\scriptsize{LHC}}} \sim
\left[10^{-7}-10^{-6}\right]$ by order of magnitude, respectively.

\section{\label{sec4}Summary}
The background magnetic field shows a significant increasing at
initial energy decreasing for intermediate and large times. Two
scenarios for initial time give rise to values close to each other
by order of magnitude in the RHIC energy domain 62.4 -- 200 GeV.
Boundary values of $\mathcal{P}$-odd correlators are estimated for
same and opposite charge combinations in $\mbox{Au+Au}$ collision
at 62.4 and 200 GeV. Preliminary STAR experimental data are
compared with predictions of chiral magnetic effect model assumed
local $\mathcal{TIP}$ violation at various centralities and
collision energies. Centrality dependencies are similar for model
results at first scenario of initial time and preliminary STAR
experimental data at 62.4 and 200 GeV for both same and opposite
charge correlators under consideration. Preliminary STAR data for
$\mbox{Au+Au}$ collisions at 62.4 GeV agree with allowed domain
for correlator values from analytic approach of chiral magnetic
effect for overall centralities reasonably. The model assumed
local $\mathcal{TIP}$ violation underestimates correlator values
for central events for $\mbox{Au+Au}$ at 200 GeV. Agreement is
some better for opposite charge correlators than that for same
charge combination. Perhaps some disagreement at highest energy
under consideration is explained by range of applicability of the
analytic approach as well as some other effects which contribute
in experimental results and not relate with local $\mathcal{TIP}$
violation hypothesis. The quantitative conclusions require, in
particular, the additional study of behaviour of background
magnetic field and more precise estimation of initial time for
phenomenological calculations and future experimental
investigation of this phenomenon.

{\em \textbf{Acknowledgments.}} I am grateful to D. E. Kharzeev
and S. A. Voloshin for discussions.

\end{document}